\documentclass[twocolumn]{aa}
\usepackage{graphics}
\usepackage{graphicx}
\usepackage{amsfonts}
\include{epsf}
\include{epstopdf}
\DeclareGraphicsExtensions{.eps}
\newcommand{\bfo}[1]{\mbox{\boldmath $#1$}}
\def\bvarphi{\mbox{\boldmath $\varphi$}}

\begin{document}
\newcommand{\beq}{\begin{equation}}
\newcommand{\eeq}{\end{equation}}
\def\la{\hbox{\raise.35ex\rlap{$<$}\lower.6ex\hbox{$\sim$}\ }}
\def\ga{\hbox{\raise.35ex\rlap{$>$}\lower.6ex\hbox{$\sim$}\ }}
\def\runit{\hat {\bf  r}}
\def\phunit{\hat {\bfo \bvarphi}}
\def\etaunit{\hat {\bfo \eta}}
\def\zunit{\hat {\bf z}}
\def\zetaunit{\hat {\bfo \zeta}}
\def\xiunit{\hat {\bfo \xi}}
\def\beq{\begin{equation}}
\def\eeq{\end{equation}}
\def\beqa{\begin{eqnarray}}
\def\eeqa{\end{eqnarray}}
\def\sub#1{_{_{#1}}}
\def\order#1{{\cal O}\left({#1}\right)}
\newcommand{\sfrac}[2]{\small \mbox{$\frac{#1}{#2}$}}
%
%
\title{{Potential vorticity dynamics in the framework of disk shallow-water theory: I. The Rossby wave instability}}

\author{O. M. Umurhan\inst{1,2}}

   \offprints{O.M. Umurhan \email{umurhan@maths.qmul.ac.uk}}

   \institute{
   School of Mathematical Sciences, Queen Mary
   University of London, London E1 4NS, U.K.\
     \and
        Astronomy Department, City College of San Francisco,
      San Francisco, CA 94112, USA\
}

\date{}

 \abstract
     {The Rossby wave instability in astrophysical disks is as a potentially important
     mechanism for driving angular momentum transport in disks. }
      {We aim to understand this instability in an approximate
      three-dimensional disk model environment which we assume to be a single
      homentropic annular layer we analyze using disk shallow-water theory.}
  {We consider the normal mode stability analysis of two kinds of radial profiles of the
  mean potential vorticity:  The first
  type is a single step and the second kind is a symmetrical step of finite width
  describing either a localized
  depression or peak of the mean potential vorticity. }
     {For single potential vorticity steps we find there is no instability.
     There is no instability
     when the symmetric step is a localized peak.  However,
     the Rossby wave instability occurs when the symmetrical step profile is
     a depression, which,
     in turn, corresponds to localized peaks in the mean enthalpy profile.
     This is in qualitative agreement with previous two-dimensional investigations
     of the instability.
       For all potential vorticity depressions, instability occurs for
       regions narrower than some maximum radial length scale.
     We interpret the instability as resulting from the interaction of at least two
     Rossby edgewaves. }
     {We identify the Rossby wave instability in the restricted three-dimensional
     framework of disk shallow water theory. Additional examinations of generalized barotropic
     flows are needed.  Viewing disk vortical instabilities from the conceptual perspective
     of interacting edgewaves can be useful. }


\titlerunning{Rossby wave instability}


  \maketitle

\section{Introduction}
The Rossby wave instability (hereafter RWI) is a promising candidate mechanism
to account  {for the observed anomalous transport of cold astrophysical disks} \footnote{i.e.,
non-magnetized disks.}.  The process involves the instability of waves
in an environment in which there are radial variations in a potential
vorticity quantity (PV hereafter).  This effect is the disk analog of the Rayleigh instability
of stably stratified barotropic shear flows that are familiar in meteorology and geophysical
fluid dynamics. 
The original works wherein the RWI was proposed for disks (Lovelace et al. 1999;
Li et al. 2000; Li et al. 2001) demonstrated the
existence and evolution of the instability in vertically integrated disk
models.   {The earliest known (to this author) identification
of the relevance of PV dynamics to astrophysical settings is found
in Lovelace \& Hohlfeld (1978) containing a study of the stability
of cold self-gravitating galactic disks
subjected to radially localized perturbations. For the galactic disk
study as well as the RWI,
instability requires
the existence of a minimum or maximum in the associated radial PV-profile of the
disk}.\footnote{Note that in Lovelace \& Hohlfeld (1978)
the associated quantity $f(r)$ corresponds roughly to the inverse of the PV.}

\par
Because of the potential importance of this mechanism for disks, it is
worthwhile to consider and understand this instability
 {in disk settings}, which are three dimensional, at least in part.
The goal of this study is to expose the machinery of this process as
clearly as possible in the context of a three-dimensional theory.
\par
Disk shallow water theory (Umurhan 2008, and hereafter DSW-theory) is a model
reduction of the disk equations that is three-dimensional but asymptotic
in the sense that azimuthal scales are much larger than the corresponding
vertical and radial scales.  It is essentially a model environment representing
vortex dynamics occurring on thin annular sections of disks over timescales which
are much longer than the local disk rotation time.  The original
motivation for this approximation was to develop a three-dimensional framework
to describe the dynamics
of elongated vortices known to emerge in two-dimensional studies
such as that reported in  Godon \& Livio (1999).
The benefit of using the DSW-theory is that it allows a transparent
analysis of vortex dynamics free of other physical processes that are
likely not to play a significant role - in particular,
free of both acoustic and gravity mode oscillations.  As such,
by means of this framework
one may develop a mechanical understanding of
disk-related vortex dynamics.\par
In Umurhan (2008), DSW-theory was applied to understand the
emergence of one form of the strato-rotational instability, which is an instability of a stably stratified Rayleigh
stable shear profile in a channel
(Yavneh 2001;
Dubrulle et al. 2005).  The no-normal flow boundary conditions
at the walls of the domain bring into existence edgewaves (Goldreich et al. 1986)
that propagate
along the walls in opposite directions with respect to one another.
However, because the edgewaves may also interact with one another across the
domain, if the separation of the walls fall within a specific allowed range, then the
phase speeds of the edgewaves can
`resonate' with each other and become unstable (Umurhan 2006; Umurhan 2008).  \par
 {This edgewave dynamical picture has been used to understand} the development
of many forms of barotropic geophysical fluid instabilities including, among others,
the original problem considered by Rayleigh (Hoskins et al. 1985; Baines \& Mitsudera 1994;
Heifetz et al. 1999)
as well as astrophysical processes such as
the Papaloizou-Pringle instability (Papaloizou \& Pringle 1984; Goldreich et al. 1985).
In the geophysical fluid dynamics context, this is frequently referred to as
the instability of counter-propagating Rossby waves (e.g., Heifetz et al. 1999).
 {The use of edgewaves as an interpretive tool has precedence
in the analysis of plasma instabilities.  For example, the diocotron effect,
which is an instability associated with finite thickness charge sheets subjected
to $E\times B$ drift, {is rationalized to arise from the interaction of
edgewaves counter-propagating along}
the surfaces of the charge sheet (Knauer 1966).}
\par
DSW-theory is an appropriate platform to examine the RWI.
Indeed, an examination of the vortex structures emerging from nonlinear calculations
of the instability (see Fig. 1 of Li et al. 2000, hereafter LFLC-2000) shows that the coherent vortex
structure is severely elongated in the azimuthal direction while being radially tightly
confined.  This structural quality is similar to the vortices
reported in Godon and Livio (1999).  The advantage of DSW-theory is that it
can capture some essence of three-dimensionality, while its disadvantage is that,
because of its construction, small azimuthal scales cannot be represented.
By contrast, the vertically integrated model environment used
in the original RWI studies are able to capture small azimuthal scales while
coming at the expense of representing three-dimensionality.\par

\begin{figure}
\begin{center}
\leavevmode
\epsfysize=8.75cm
\epsfbox{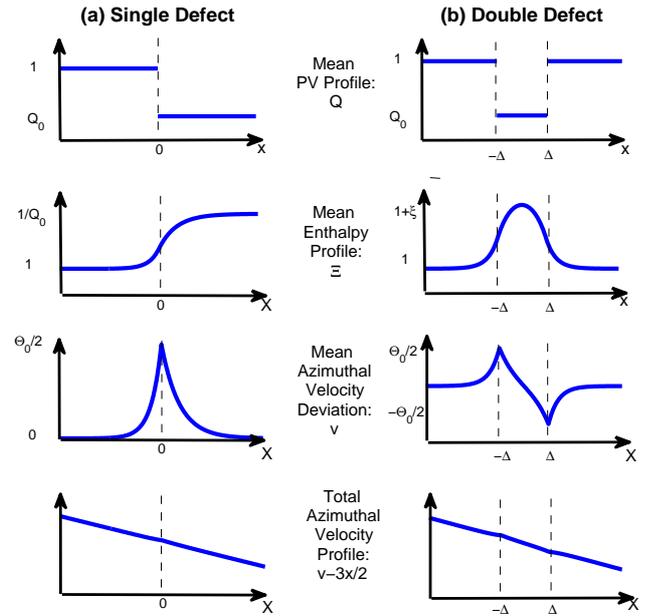}
\end{center}
\caption{Sketches of profile plots considered in this work.  (a) Single defect/jump located
at $x=0$.  (b) Double defect (symmetric step) with jumps located at $x=\pm\Delta$.  Double
defect depicted is a depression whose mean (midplane) enthalpy profile corresponds to a peak.}
\label{profiles}
\end{figure}

In this study, the RWI is examined in the framework of DSW-theory.
The stability of two analytically tractable mean PV-profiles is studied: (i)
a single defect, or `single-step', and (ii) a double defect, or `symmetric-step'
(see Fig. \ref{profiles}).
  As in the case of the stratorotational instability, the RWI
is shown to result from the interaction of Rossby edgewaves
that propagate along the locations of the steps of the mean PV-profile.  However,
the instability manifests itself only in the symmetric step case.
In Section 2, the equations of motion
and their perturbations are presented.  In Section 3, the equations are analyzed for
two kinds of mean potential vorticity profiles including the one just described.
Analytic solutions are determined
throughout by considering the quasi-linear problem posed by considering the special
case $\gamma = 3$, where $\gamma$ is the ratio of specific heats.
Section 4 concludes with comments and a discussion about
recent three-dimensional RWI calculations (Meheut et al. 2010).

\section{Equations and potential vorticity conservation}
The asymptotic equations upon which DSW-theory is based, describing the  evolution of annular
disk sections, was developed in Umurhan (2008).  In that study, the equations
of motion for an annular section in which the vertical/radial extent ${\cal L}$ is much smaller
that the azimuthal scale $R$ was developed as an asymptotic series expansion in
powers of $\epsilon \equiv {\cal L}/R$.  The timescales of
the dynamics (${\cal T}$) is comparatively long with respect 
to the local rotation time ($1/\Omega_0$) of the annular disk centered
on $R_0$ i.e., ${\cal T} \sim 1/(\epsilon \Omega_0)$.    To ensure a non-trivial
asymptotic balance, the radial and vertical velocities viewed in the reference
 frame of the annulus rotating with $\Omega_0$ must scale as $\epsilon c_s$ where
$c_s$ is the sound speed of the gas, while the corresponding deviations of
the azimuthal velocities must scale as $c_s$.  This disparity in aspect and velocity ratios 
is an asymptotic description of azimuthally elongated slowly overturning
vortices.  Furthermore,  the reduced equations
of motion in the annular disk section possess quasi-geostrophic characteristics
familiar in meteorological
studies.  \par
Single constant entropy layers are considered when the equation
of state is polytropic, i.e., it is assumed that $P = K\rho^\gamma$ everywhere at all times.
This corresponds to a situation in which the specific entropy inside, defined by $S = C_v\ln(P/\rho^\gamma)
=C_v\ln K$, is everywhere a constant.  This homentropic layer is considered in the following.
If there are radial variations of the mean height  of the disk ($h$), then
the {\em vertically integrated} entropy of the disk also exhibits radial variations even though
the specific entropy is constant.
This point is made to keep in mind how the equations analyzed here
compares to the vertically integrated two-dimensional equations studied in
Li et al. (2000).
\par
After non-dimensionalizing all the quantities according to the above,
the leading order equations
describing the evolution of varicose disturbances
of a single constant specific entropy layer are
\beqa
v &=& \sfrac{1}{2}\partial_x \Xi, \label{QG_v} \\
\frac{D v}{Dt} + (2-q) u &=& - \partial_y \Xi, \label{v_eqn} \\
\frac{D\Xi}{Dt} &=& 2\Xi \Omega_z, \label{H_eqn} \\
\partial_x u + \partial_y v + \frac{\gamma + 1}{\gamma - 1} \Omega_z &=& 0,
\label{eqn_of_state}
\eeqa
where $x,y$ represent the non-dimensional radial and azimuthal coordinates
and $t$ is the non-dimensionalized time (Umurhan, 2008).
In this construction,
all quantities appearing are functions of these two coordinates and time.
The radial velocity is $u$.
The total time derivative
is
\[
\frac{D}{Dt} \rightarrow \partial_t + (v - qx)\partial_y + u\partial_x.
\]
The background shear $=-q x {\hat {\bf y}}$, is linear in the radial coordinate.
For a Keplerian disk, $q=3/2$.
Thus, the azimuthal velocity $v$ appearing here
is understood to be a deviation from this basic background state as viewed in the
local rotating frame.
The quantity $\Xi$
may be understood to be the enthalpy of the disk midplane and is given by
\beq
\Xi = \sfrac{1}{2} h^2,
\eeq
where $h=h(x,y,t)$ corresponds to the height of the disk measured from the disk midplane.
 The
height $h$ is the position at which 
the enthalpy goes to zero in the constant (specific) entropy environment.
Although $\Xi$ is technically the \emph{midplane enthalpy}, it is
referred to hereafter as simply the {\em enthalpy} with full understanding that it is
really its value at the midplane.\par
Equation (\ref{QG_v}) is the radial momentum balance equation but in this asymptotic limit it says
that all perturbations are radially geostrophic.  Equation (\ref{v_eqn}) is the azimuthal momentum balance
and, unlike the radial version, no geostrophic balance is implied.  Equations (\ref{H_eqn})
and (\ref{eqn_of_state}) represent, respectively, the motion of the surface $h$ and an
equation of state for the gas (see below).  In the derivation contained in
Umurhan (2008), these two equations result from (i) exploiting the hydrostasy of
perturbations at all times and (ii) an analysis of the energy equation, which
makes use of the linear
independence of powers of $z$.\par
The disk vertical velocity, though not explicitly appearing,
is odd symmetric with respect to the disk midplane because of the assumed varicose symmetry of the disk response.  In the development of the DSW-theory its functional form is shown to be
$w = z\Omega_z(x,y,t)$.
In the reduction leading to these equations only
$\partial_z w$ appears as well as the explicit evaluation of the velocity
of the layer at the height $z=h$. Additional details are found in Umurhan (2008).
\par
The combination of these equations of motion leads to the conservation of potential vorticity $Q$
given by
\beq
\frac{DQ}{Dt} = 0, \qquad Q \equiv \frac{2(2-q) + \partial_x^2 \Xi}{\Xi^{\frac{1}{2}\frac{\gamma+1}{\gamma-1}}}.
\eeq
The next section considers the fate of normal modes for potential vorticity (or `PV' for short)
profiles in which
there is either a single or a double defect in the steady PV profile $Q = \bar Q(x)$.
\section{Analytical study}
The analysis in this work considers the fate of disturbances where $\gamma = 3$.  This
is chosen because the resulting equations, both steady and normal mode disturances, may
be analyzed without intensive use of numerical methods. General more realistic values of
$\gamma$ require numerical evaluation of both the steady and time-depenedent state and
this will be reserved for a future study.  For $\gamma = 3$, the PV expression takes on
the quasi-linear form
\beq
Q = \frac{2(2-q) + \partial_x^2 \Xi}{\Xi}.
\eeq
The following sections consider the two analytically tractable defect profiles and their
normal mode responses.
\subsection{Single defect}
In this study, the term {\em defect} is used to indicate places where steps
occur in the mean PV-profile.  The following simple single defect,
\beq
\bar Q = \left \{ \begin{array}{cr}
1, & x<0, \\
Q_0 & x> 0,
\end{array}
\right.
\eeq
is studied.
The value of $Q_0$ is constrained to be greater than zero.   Solutions are sought that
are bounded as $x\rightarrow \pm \infty$.
 This means solving for the mean
enthalpy $\bar \Xi$
\beq
\partial_x^2\bar \Xi - \bar Q \bar \Xi = -1 ,
\eeq
on either side of $x=0$
(and where $q = 3/2$ has been explicitly used).
The particular boundary conditions for the profiles at infinity
are (i) $\bar\Xi \rightarrow 1$ as $x\rightarrow -\infty$ and (ii)
 $\bar\Xi \rightarrow 1/Q_0$ as $x\rightarrow \infty$.
 At the interface $x=0$, it is also required that $\bar \Xi$
 and its radial gradient $\partial_x \bar \Xi$ are matched, the latter
 being identical to requiring the matching of the
 azimuthal velocities $\bar v$ from both sides.
The solution for the mean enthalphy is therefore
\beq
\bar \Xi =
 \left \{ \begin{array}{cr}
1 + \Theta_0 e^{x}, & x<0, \\
1/Q_0-\Theta_0/\sqrt Q_0 e^{-\sqrt Q_0 x} & x> 0,
\end{array}
\right.\eeq
where
\beq
 \Theta_0 = \left(\frac{1-Q_0}{Q_0}\right)\frac{\sqrt{Q_0}}{1 + \sqrt{Q_0}}.
\eeq
This result may be read to mean that the mean height $\bar h$ decreases when
$Q_0 > 1$ and increases when $0<Q_0<1$.
It follows from Eq. (\ref{QG_v}) that
\beq
\bar v = \sfrac{1}{2}
\left\{ \begin{array}{cr}
\Theta_0 e^{x}, & x<0, \\
\Theta_0 e^{-\sqrt Q_0 x}, & x> 0,
\end{array}
\right.
\eeq
which means that the deviation velocity at $x=0$ is given  by
$\bar v_0 = \Theta_0/2$.   As $x\rightarrow \pm\infty$,
the mean velocity deviation goes to zero.
The mean states developed here are qualitatively sketched in Fig. \ref{profiles}.
\par
Infinitesimal perturbations of this steady state are
introduced by writing for  each dependent quantity
$F \rightarrow \bar F + F'$, where $F'(x,y,t)$ is small
compared to the basic state $\bar F$.  This means that
the potential vorticity equation is
\beq
\Bigl[\partial_t + (\bar v - 3x/2)\partial_y\Bigr] Q'
+ u' \frac{d\bar Q}{dx} = 0,
\eeq
where
$Q'$ is given as
\[
Q' = \frac{\partial_x^2 \Xi' - \bar Q \Xi'}{\bar \Xi}.
\]
Because $d\bar Q/dx$ entails a delta-function at $x=0$,
special well-known treatments must be utilized to solve these equations
(for example, see Drazin \& Reid 1982).
For this type of problem, the procedure is detailed, for example, in Umurhan (2008),
and may be summarized as follows: (i) solve
$Q' = 0 $ on either side of the defect, (ii) match $\Xi'$
coming in from both sides, and (iii) make sure that
the normal velocities $u'$ are matched across the boundary.
The last of these may be straightforwardly enforced by
inspecting the linearized version of Eq. (\ref{v_eqn}),
\beq
\Bigl(\partial_t + (\bar v - qx)\partial_y\Bigr)\partial_x \Xi'
+\Bigl(4-2q + \partial_x^2\bar \Xi\Bigr) u' = - 2\partial_y \Xi',
\eeq
and requiring that the two solutions coming in from
either side of $x=0$ satisfy the requirement
that the normal velocities $u'$ are the same there.
\par
Normal mode solutions of the form $\Xi' = \hat \Xi(x)\exp{i\alpha(y-ct)}$ are assumed.
Solutions with $c>0$ correspond to waves that propagate in the forward $y$ direction
and, correspondingly, in the negative $y$ direction for $c<0$.
It follows that the solution for $\Xi'$, away from the two boundaries,
satisfying the condition that their amplitudes are equal at $x=0$
and,  $\hat \Xi \rightarrow 0$ as $x\rightarrow \pm \infty$, is
\beq
\hat \Xi = A_0
\left\{ \begin{array}{cr}
 e^{x}, & x<0, \\
e^{-\sqrt Q_0 x}, & x> 0,
\end{array}
\right.
\eeq
where $A_0$ is an arbitrary constant.  The requirement that the
normal perturbation velocities,
$u'$, match at the defect location means
ensuring that
\beq
\left[
\frac{(\bar v_0-c-3x/2)\partial_x\hat\Xi + 2\hat\Xi}{1 + \partial_x^2\bar \Xi}
\right]^{x\rightarrow 0^+}_{x\rightarrow 0^-} = 0.
\eeq
Satisfaction of this condition leads to an eigenvalue condition on
the wavespeed $c$ which, after some algebra, reduces to the simple
form
\beq
c - \bar v_0 = -2\Theta_0 = -4\bar v_0.
\label{single_wavespeed}
\eeq
For $\bar v_0$ given above, it follows that
that $c = -3\bar v_0$.  The wave here is a Rossby edgewave phenomenon
(e.g. Baines \ Mitsudera 1994)
and the eigenvalue determined here describes the speed with which
the edgewave propagates.  If the defect were placed at some
position other than $x=0$, then the observed speed of the edgewave
would be Doppler-shifted according to the Keplerian velocity at
the position of the defect.  Thus, if the defect were instead at
$x=x_0$, the wavespeed would then be given by $c = -3x_0/2 - 3\bar v_0$.

\subsection{Double defect: The symmetric step}
{This subsection is an analysis of a symmetrical defect pair}.  The previous
section demonstrates how a single defect produces a single
Rossby edgewave along the defect.  For two defects, therefore, two
Rossby edgewaves are expected and it is under these circumstances that
instability may emerge according to the principles of
interacting waves (Hayashi \& Young 1985;
Baines \& Mitsudera 1994;  Heifetz et al. 1999).
\par
The mean PV-profile appropriate for a double symmetric defect
is
\beq
\bar Q =
\left\{ \begin{array}{cc}
 1, & x<-\Delta, \\
Q_0, & -\Delta< x< \Delta, \\
1, & x>\Delta,
\end{array}
\right.
\eeq
where $\Delta > 0$.
The edges of the defect are located at $x=\pm \Delta$ and separated by
a distance $2\Delta$, where $\Delta$ is the region's half-width.
Following the same procedures as outlined in
the previous section
and requiring that $\bar \Xi \rightarrow 1$ as $x\rightarrow \pm \infty$,
the mean enthalpy is found to be
\beq
\bar \Xi =
\left\{ \begin{array}{cc}
 \Theta_0 e^{x+\Delta} + 1, & x<-\Delta, \\
  & \\
\frac{1}{Q_0} - \frac{\Theta_0}{\sqrt{Q_0}}
\frac{\cosh{\sqrt {Q_0}x}}{\sinh{\sqrt{Q_0}\Delta}}, & -\Delta <x< \Delta, \\
 & \\
 \Theta_0 e^{-x+\Delta} + 1, & x>\Delta.
\end{array}
\right.
\eeq
The value of $\Theta_0$ in this case is
\beq
\Theta_0 = \frac{1-Q_0}{Q_0}\cdot
\frac{\sqrt{Q_0}\tanh{\sqrt{Q_0} \Delta}}{1+\sqrt{Q_0}\tanh{\sqrt{Q_0} \Delta}}.
\eeq
It means that for $0<Q_0<1$, the enthalpy deviation is a \emph{peak} or \emph{bump}, while for $Q_0>1$
the enthalpy structure is a \emph{depression}.
The corresponding mean velocity deviations are
\beq
\bar v =\frac{1}{2}
\left\{ \begin{array}{cc}
 \Theta_0 e^{x+\Delta} , & x<-\Delta, \\
  & \\
 - \Theta_0
\frac{\sinh{\sqrt {Q_0}x}}{\sinh{\sqrt{Q_0}\Delta}}, & -\Delta <x< \Delta, \\
 & \\
 -\Theta_0 e^{-x+\Delta}, & x>\Delta.
\end{array}
\right.
\eeq
Defining $\bar v({\pm\Delta}) \equiv \bar v (x=\pm \Delta)$, it follows
that at the defects $\bar v({\pm\Delta}) = \mp\Theta_0/2$. It is useful
to define $\xi$, i.e., the ratio of the height of the peak mean enthalpy to its
value as $x\rightarrow \pm \infty$,
\beq
\xi \equiv \frac{{\rm{max}}{(\bar \Xi})}{\bar\Xi(\pm\infty)}
=
\frac{1}{Q_0} - \frac{\Theta_0}{\sqrt{Q_0}}
\frac{1}{\sinh{\sqrt{Q_0}\Delta}}.
\label{definition_xi}
\eeq
A qualitative sketch of the mean states are depicted in Fig. \ref{profiles}.

\begin{figure}
\begin{center}
\leavevmode
\epsfysize=4.7cm
\hskip -.5cm
\epsfbox{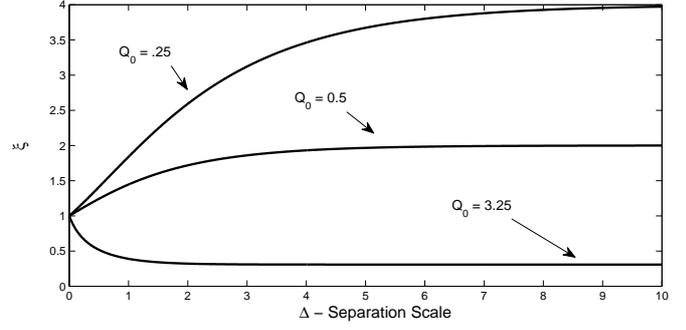}
\end{center}
\caption{Plot of $\xi$, the ratio of the minimum or maximum of the mean enthalpy
 to its value as $x\rightarrow \pm\infty$ (symmetric step profile only).  Three
values of $Q_0$ are shown: $Q_0 = 0.25,0.5$ (depression) and $Q_0 = 3.25$ (peak).}
\label{xi_plots}
\end{figure}
\par
\begin{figure}
\begin{center}
\leavevmode
\epsfysize=7.25cm
\epsfbox{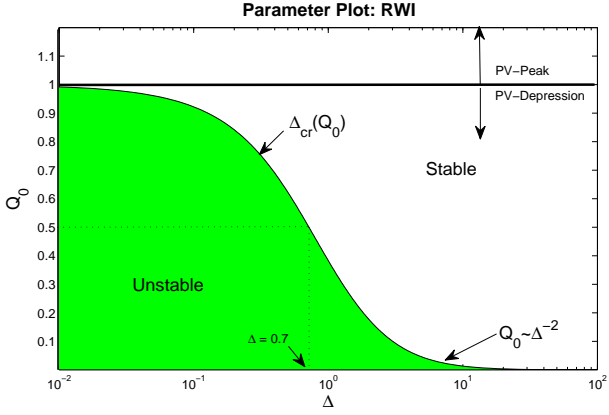}
\end{center}
\caption{Stability boundaries for the double defect problem in the $\Delta-Q_0$ plane.
Unstable region depicted by shaded zone.  Instability occurs for all values of enthalpy bumps
(i.e., PV-depressions: $0<Q_0<1$). No instability expected for enthalpy depressions.
The boundary $\Delta = \Delta_{{\rm cr}}(Q_0)$ separates stable and unstable regions.
Critical values of $Q_0$ corresponding to instability goes like $\Delta^{-2}$ for $\Delta \gg 1$.}
\label{Figure1}
\end{figure}
\begin{figure}
\begin{center}
\leavevmode
\epsfysize=7.5cm
\epsfbox{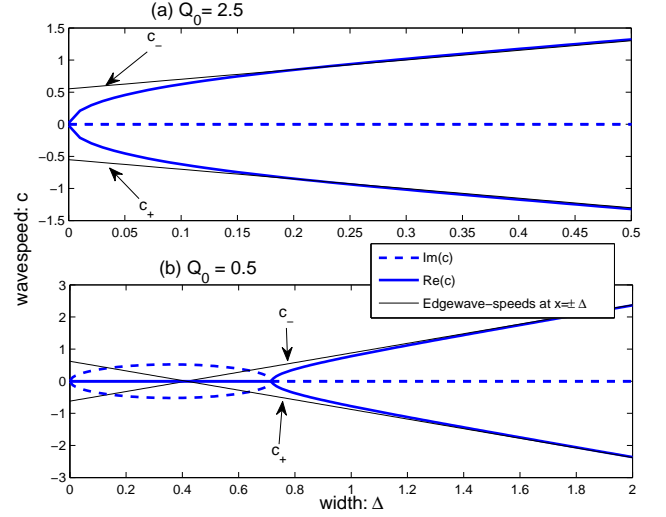}
\end{center}
\caption{wavespeeds as a function of $\Delta$ for symmetric step profile: (a) $Q_0 = 2.5$, (b) $Q_0 = 0.5$. Also shown are the speeds $c_\pm$ (discussed in the text, Sect. 4)
corresponding to the edgewaves that would propagate along each defect \emph{as
if each defect were treated
in isolation}.
Instability occurs in (b) only when these two wavespeeds are nearly equal,
which is consistent with the Hayashi-Young criterion for wave instability.}
\label{growth_vs_Delta}
\end{figure}

Because there are two defects, individual edgewaves
propagate along each of them, which interact with
each other as well.  A normal-mode analysis, as
executed in the previous section, is performed.  The details
are presented in the Appendix.  The resulting wavespeed $c$
satisfies the relationship
\beq
c^2 = {\cal E},
\label{double_defect_wavespeed_relationship}
\eeq
where
\beqa
& & {\cal E}(Q_0,\Delta) =
\frac{3(Q_0-1)}{4Q_0}\biggl\{
1 + \frac{4\beta}{\sqrt {Q_0}}
 \nonumber \\
& & -\frac{3\beta^2}{1-Q_0} - \frac{1}{(\cosh\beta + \sqrt{Q_0}\sinh \beta)^2} \nonumber \\
& & +\frac{4\left(\beta+{\sqrt{Q_0}}\right)\sinh\beta
\left[(Q_0-2)\cosh\beta - \sqrt{Q_0}\sinh\beta\right]}
{2Q_0\cosh {2\beta}  + \sqrt{Q_0}(1+Q_0)\sinh{2\beta}}
\biggr\}, \nonumber
\eeqa
and $\beta \equiv \sqrt{Q_0}\Delta$ is defined for convenience.
The results of this are plotted in Fig. \ref{Figure1}.  The first clear result
is that instability (${\cal E} = 0$) may only occur for enthalpy bumps, which is
to say that
instability occurs
for PV-depressions, $0<Q_0<1$.  The boundary between stable and unstable
parameter regimes as a function of $Q_0$ is given by $\Delta_{{\rm cr}}(Q_0)$.
Thus, for PV-depressions, instability occurs for $0<\Delta <\Delta_{{\rm cr}}$.
For very small but non-zero values of $Q_0$, the results show that
$\Delta_{{\rm cr}} \sim Q_0^{-1/2}$.  Defining instead $Q_{{\rm cr}}$ as
the marginally stable value of the PV-depression then its asymptotic
form has the functional dependence $\sim \Delta^{-2}$ for $\Delta \gg 1$.
Figure \ref{growth_vs_Delta} shows plots of the wavespeeds as a function
of $\Delta$ for representative values of a PV depression and peak.

\section{Comments and discussion}
It has been shown that the RWI is recovered in the framework
of DSW-theory.  The RWI emerges when the idealized
symmetric-step mean (radial) potential-vorticity profile is a depression (or equivalently,
when the mean radial enthalpy profile is a peak).  { However,} single stepped mean PV profiles
do not support the instability.  The RWI is interpreted as being caused by the interaction of two
Rossby-edgewaves propagating azimuthally along each defect of the symmetric step PV-profile.
These types of dynamical processes are a
generic feature of many vortical dynamics in geophysical flows.
Below is a collection of comments about these results including comparisons 
between the results of this work and the original RWI studies:
\par
\bigskip
1. A weakness of the analysis performed here is that an unrealistic
value of the ratio of specific heats was used: a value of $\gamma = 3$
places the fluid in a thermodynamic regime somewhere between a normal gas
and an incompressible fluid (the latter being $\gamma \rightarrow \infty$, Salby, 1989).
The main reason for this choice
was to develop the analysis analytically as most other values of $\gamma$
require numerical methods to study both the steady states and the perturbation evolution.
Among other things, developing analytical solutions are important for benchmarking future analyses for
more reasonable values of $\gamma$.  But most of all, the analytical solutions offer
the most transparency in interpreting the results.
Preliminary calculations for realistic values of $\gamma$ show that the main mechanical interpretations
and qualitative solutions, as well as conditions for onset, are not considerably altered.  
This is the subject
of a study in preparation.\par
2. The limiting form of the DSW-theory used to study the RWI considers the response of a
single homentropic disk layer.  In this picture, the mean vertical height
of the disk is a function of the radial coordinate, i.e., $\bar h = \bar h(x)$.
Instability only occurs when the enthalpy of the disk layer (which is $\propto \bar h^2$)
has a peaked structure.  In LFLC-2000, the RWI is demonstrated in a vertically integrated model of
a disk.  Similar in quality to the results here, instability occurs when there is a
peak in the mean radial pressure profile (or, relatedly, the surface density profile).  In contrast, however, instability occurs
for constant specific entropy profiles.  This might imply that these
results conflict with one another.  \par
Because the disk model considered in LFLC-2000 is vertically integrated, all
of the fluid quantities considered there are also understood to be vertically integrated, 
including the entropy.  In the DSW-theory framework studied here, despite
the specific entropy being constant throughout the layer, its vertical integration
is not constant if the mean height of the disk is not uniform.  That is to say
if one defines
\[
{\cal S} = \int^{-\bar h}_{\bar h} S dz = 2\bar h C_v \ln K,
\]
as the vertically integrated specific entropy of the disk, then it is self-evident
that ${\cal S}$ is not uniform if $\bar h$ varies with radius $x$.  As such, the results
of this study should be compared to the non-homentropic incarnations of the RWI
 discussed in the original studies, for example, as in Li et al. (2001).
 \par
3. With respect to the last point, the version of the RWI uncovered in the analysis
of the symmetric step is qualitatively similar to the homentropic Gaussian bump profile
(`HGB') considered in LFLC-2000, where HGB describes the
quality of the radial dependence of the mean pressure.
However, as stated above, the symmetric step profile
studied in this work is effectively non-homentropic when viewed in terms of
the vertically integrated disk model considered in Li et al. (2000).
The symmetric step profile might, therefore, most closely correspond to a {\it
non-homentropic Gaussian bump profile} (`NGB') in the framework of the vertically
integrated model of LFLC-2000.  A NGB profile was examined in the nonlinear
regime in  Li et al. (2001), and, in a qualitative sense,
its evolution was not found to differ considerably.
\par
One may, therefore, reasonably compare the conditions in which the RWI
arises in the HGB setting to the instability arising here for the
symmetric-step profile.   In the analysis presented in LFLC-2000, it was shown
that for the HGB profile a minimum amplitude in the pressure bump was required
for instability to occur.   It was indicated in Fig. 9. of that study
that the minimum variation required
in the associated surface mass density for instability to occur, on a
radial scale similar to the thickness of the disk,
must be between approximately 10\% to 20\% (LFLC-2000).
According to this study,
for any given value of $0 <Q_0 < 1$, instability occurs
for $0<\Delta<\Delta_{{\rm cr}}$.  This means that for a given
peak value of the midplane enthalpy there is a maximum width size beyond
which instability ceases.  In this sense, the results here imply
that there is no minimum required amplitude of the
midplane enthalpy peak (and, consequently, the corresponding
associated surface density peak)
for instability to occur. \par
This does not necessarily
mean that these results disagree with each other. Indeed
in Fig. 3, it can be seen that for fixed values of $\Delta$,
there is a value of $Q_0$ smaller than some critical value 
for the system to be unstable.  For example, for
$\Delta = 0.7$, instability occurs for $Q_0 < 0.5$.  According
to Fig. 2, this means that for
these parameter values the minimum midplane enthalpy peak variation
over the mean must be at least 25\% for instability to occur.
\par
This still leaves open the question about the origin of the initial equilibrium.
In the general scenario proposed in LFLC-2000, matter is accumulated
in a piecemeal way, either by means of the coupling to large-scale magnetic field
processes or due to small-scale turbulence (possibility MHD driven).  These
possibilities are not disputed here.  Instead attention should be paid
to the way in which matter is accumulated.  Reference to Fig. 3 best
illustrates this point.   {The way that matter accumulates
 depends upn the path within the $Q_0$-$\Delta$ plane along which the accumulation
process takes place.}  For example, if the process involves an increase in $\Delta$
(the region size) for fixed values of $0<Q_0<1$, then instability will occur
before any sizeable enthalpy peak is reached
because any such path (starting from $\Delta = 0$) is unstable to the RWI.
On the other hand, if the initial equilibrium $Q_0$ is constructed on a path
of constant $\Delta$, then the system may be brought into the unstable regime
with a finite large amplitude of the enthalpy peak variation.
Thus, the way in which the equilibrium is constructed is crucial to assessing
how and on what scale the instability will operate.  This point deserves further attention
in the future.
\par
4. The single step profile in the mean PV
considered in this analysis shows no instability.  In contrast,
the single jump profiles studied in LFLC-2000 do show
the emergence of instability.  The reason for this difference
is not entirely clear.  One possibility is that the single
step profile considered here
may not be a faithful analog of the single jump
profile studied in LFLC-2000.  Indeed, in the generalized picture of
edgewave dynamics of Baines \& Mitsudera (1994), one needs
a minimum of two edgewaves for instability to happen.
The single-defect model of this study can only support a single edgewave
and, therefore, no instability is possible.  A corresponding
analog representation of the single jump profile considered in LFLC-2000
probably needs to support at least two edgewaves in order to see
the instability expressed.  This is, however, conjecture at this stage
and more analysis is needed.

5.  {Instability can be best rationalized in terms of the interaction
of the two individual edgewaves propagating along each defect
of the symmetric step profile}.  This picture is consistent with the
Hayashi-Young criterion for wave instability (Hayashi \& Young 1987),
which can be summarized as follows:  instability of two
waves separated by an evanescent layer occurs when the two waves
(i) propagate opposite to each other, (ii) have almost
the same Doppler-shifted frequency, and (iii) can
interact with one another.  Interaction
in this case means that the edgewave at one defect contributes to the perturbation
velocity profile at the other defect.\par
All of these conditions for wave instability
are satisfied in the problem encountered here.
In particular,
one can view the Rossby-edgewaves propagating along each
defect as though they were disturbances
of \emph{isolated} defects.  Thus,
with respect
to the wavespeed calculation worked out for single isolated steps in PV
(Sect. 3.1), one can write down the wavespeeds associated
with each defect as
\[
c_- = \frac{3}{2}\Delta - \frac{5}{2}\Theta_0
 = \frac{3}{2}\Delta - \frac{5}{2}\frac{1-Q_0}{Q_0}
\cdot \frac{\sqrt{Q_0}}{1+\sqrt{Q_0}},
\]
for the wave at $x=-\Delta$ and
\[
c_+ = -\frac{3}{2}\Delta + \frac{5}{2}\Theta_0
 = -\frac{3}{2}\Delta + \frac{5}{2}\frac{1-Q_0}{Q_0}
\cdot \frac{\sqrt{Q_0}}{1+\sqrt{Q_0}},
\]
at $x=\Delta$.  The wavespeeds $c_{\pm}$ and
the values of $c$ appropriate for the double-defect problem
are plotted in Fig. \ref{growth_vs_Delta}.
\footnote{The single wavespeed formula
quoted above for $c_-$
is written in terms of the formula given in
Eq. (\ref{single_wavespeed}) augmented by
the Doppler-shift consistent with the the local Keplerian
speed at $x=-\Delta$.  The isolated edge wavespeed $c_+$ is
correspondingly written taking into consideration
the symmetry of the mean PV-profile as viewed at $x=\Delta$, which
is the opposite of that at $x=-\Delta$. }
It can be seen in Fig. \ref{growth_vs_Delta}b that instability occurs when
the possibility exists for the two isolated waves to have nearly
the same (and opposite) wavespeeds.  In the rest frame $x=0$,
this occurs when each of the wavespeeds
are nearly equal to zero.  In Fig. \ref{growth_vs_Delta}a,
where instability does not occur, the wavespeeds are not nearly
equal enough (they are not close to zero), hence instability cannot occur.  \par
In this particular double-defect setup it follows that one can roughly predict
the onset of instability to be when the isolated wavespeeds are
nearly zero in the reference frame of an observer at $x=0$.  Using the
formula for $c_\pm$ given above, $c_\pm \approx 0$ happens very nearly when
\[
\Delta \approx \frac{5}{3}\frac{1-Q_0}{Q_0}
\cdot \frac{\sqrt{Q_0}}{1+\sqrt{Q_0}}.
\]
Clearly this can never happen when $Q_0 > 1$ as $\Delta$ must be positive.  Furthermore, when $Q_0 \ll 1$
this rough criterion means that the critical value of the separation scale
is approximately
\[
\Delta_{{\rm cr}} \sim Q_0^{-1/2},
\]
which is the same trend observed in Fig. \ref{Figure1}.  \par
The importance
of this perspective is that one may approximately assess whether or not
a more complicated looking PV-profile can become unstable to the RWI instability
by appealing to this conceptually simple mechanistic view.  For example, if a more sophisticated
PV-profile were treated as if it were a series of PV-steps, then an approximate
prediction for the onset of instability would only require calculating
edgewave speeds along each defect taken as though each were in isolation.
Onset conditions would be predicted for values of the parameters in which
any two of these wavespeeds are nearly equal.
\par
6.  The asymptotic limit that the DSW-theory exploits is one in which
the azimuthal length scales are much longer than the radial ones.  In this
 limit there is no information as to how wavespeeds depend
upon the horizontal wavenumber $\alpha$.  In the theory developed here
the wavespeed $c$ is independent of $\alpha$. However, it is known that in
the vertically integrated configurations investigated in the original RWI studies
there is a clear dependence of $c$ with $\alpha$.  \par
On more concrete terms, in the numerical solution shown in LFLC-2000 there is
obviously
a fastest growing azimuthal wavenumber of the instability, which
leads to the final steady pattern state reached by the system.  In contrast, in the
approximate study done here for the symmetric-step profile,
once stability is breached all wavenumbers become
unstable and there is no high-wavenumber cut off.  The growth rates,
$\sigma = \alpha {\rm Im}(c)$, obviously increase without bound
in proportion to  $\alpha$ - signaling the breakdown of the theory.  \par
To correct this issue, the DSW-theory would
have to be appropriately extended by allowing for
vertical structure in the horizontal velocities (i.e., inclusion of the
dependence $\sim z^2$) and some deviations from the linear dependenc
of the vertical velocity
with the $z$ coordinate (i.e., an additional $\sim z^3$ dependency).
 The resulting
next order correction to the
wavespeed in the extended DSW-theory
will show that the general dispersion relationship takes the form
\[
c^2 = {\cal E} + {\cal E}^{(2)}\alpha^2 ,
\]
where ${\cal E}^{(2)}$ is a function of the parameters $Q_0$ and $\Delta$.
Preliminary calculations show that
${\cal E}^{(2)}$ is greater than zero, which means
there is a high-wavenumber cutoff beyond which there would be
no instability in the extended theory.  The DSW-theory may be extended to higher orders in
$\epsilon$ by asymptotically developing
corrections for $c$ near values of ${\cal E} \approx 0$, which occur
for values of $\Delta$ near $\Delta_{{\rm cr}}$.  Further investigation is needed
in this regard.

7.  Three-dimensional numerical simulations of the RWI were announced during the
preparation of this manuscript.  { Meheut et al. (2010) report on three-dimensional
simulations of a constant entropy disk with midplane symmetric perturbations
with the main goal of following the evolution of a Gaussian
bump profile in the initial pressure field.}
Notable features of the results show that there is a significant amount of
three-dimensional structure in the horizontal velocities as well as deviations from
a linear $z$ dependence of the vertical velocities.  The resultant structures show
interesting and complex three-dimensional structure in the streamlines of the flow.  In contrast,
in the asymptotic three-dimensional theory used here the horizontal velocities have
no vertical variation and the vertical velocity is linear in $z$.  Thus, the asymptotic
strategy used here is far from matching the overall structure contained in the results
of these particular simulations.
The DSW-theory (or an analogous approach) may be extended to capture
the quality of the dynamics contained in these numerical simulations much in line
with the previous comments made in this regard.  Moreover, it is important to understand how the
three-dimensional structure emerges in these simulations, i.e., whether it arises from other secondary
 {barotropic} processes or if they are dynamics that are tied to the primary
quasi-two dimensional dynamics driven by the primary instability.  Rationalization
of the three-dimensional dynamics in terms of the Rossby edgewave perspective
utilized in this work  {could be a} natural first step to this end.


\section{Acknowledgements}
The author thanks Yiannis Tsapras, Oliver Gressel, Joe Barranco, and James Cho for
discussions relating to this work.  The author also thanks the anonymous referee
for pointing out the analogy between the diocotron effect and the RWI.  The author is indebted to
the Kavli Institute of Theoretical Physics and the program Exoplanets Rising: Astronomy and Planetary Science at the Crossroads,
where portions of this work were performed.

\appendix
\section{Double-defect normal-mode calculation}\label{DoubleDefect}
Following the procedure performed for the single defect, the equation
$Q' = 0$ is solved separately in the three zones subject to the
condition that $\hat\Xi \rightarrow 0$ as $x\rightarrow \pm \infty$.
For $x<-\Delta$ and $x>\Delta$, the equation to be solved is
\beq
\partial_x^2 \hat \Xi - \hat \Xi = 0,
\eeq
while for $-\Delta < x < \Delta$, it is
\beq
\partial_x^2 \hat \Xi - Q_0 \hat \Xi = 0.
\eeq
The solution
for $\hat \Xi$ in each of the three zones is
\beq
\bar \Xi =
\left\{ \begin{array}{cc}
 A_- e^{x+\Delta} + 1, & x<-\Delta, \\
  & \\
\sfrac{1}{2}(A_++A_-)
\frac{\cosh{\sqrt {Q_0}x}}{\cosh{\sqrt{Q_0}\Delta}},
&   \\
\ \ \ \ \ +\sfrac{1}{2}(A_+-A_-)
\frac{\sinh{\sqrt {Q_0}x}}{\sinh{\sqrt{Q_0}\Delta}},
& -\Delta <x< \Delta, \\
 & \\
 A_+ e^{-x+\Delta} + 1, & x>\Delta,
\end{array}
\right.
\eeq
The solution as presented satisfies the continuity of $\hat\Xi$
at the boundaries $x=\pm\Delta$.  The remaining conditions are
the continuity of the normal velocities at the location of
the defects, which amounts to simultaneously satisfying
\beq
\left[
\frac{(\bar v(\Delta) - 3\Delta/2 -c)\partial_x\hat\Xi + 2\hat\Xi}{1 + \partial_x^2\bar \Xi}
\right]^{x\rightarrow \Delta^+}_{x\rightarrow \Delta^-} = 0,
\eeq
and
\beq
\left[
\frac{(\bar v(-\Delta) + 3\Delta/2 -c)\partial_x\hat\Xi + 2\hat\Xi}{1 + \partial_x^2\bar \Xi}
\right]^{x\rightarrow -\Delta^+}_{x\rightarrow -\Delta^-} = 0.
\eeq
These two equations result in the matrix equation
\[
{\bf M} \bf A = 0,
\]
where ${\bf M}$ is a $2\times2$ matrix and ${\bf A} = (A_+, A_-)^\texttt{T}$.  Nontrivial
satisfaction
of this matrix equation requires enforcing $\det{\bf M} = 0$.  This results in the wavespeed
relationship found in Eq. (\ref{double_defect_wavespeed_relationship}).
\end{document}